\documentclass[journal=jacsat,manuscript=article]{achemso}
\setkeys{acs}{articletitle = true,maxauthors = 1000}

\usepackage{chemformula} 
\usepackage[T1]{fontenc} 

\usepackage{xspace}
\usepackage{color}

\newcommand{\VSe}{VSe\textsubscript{2}\xspace}%
\newcommand{\MoS}{MoS\textsubscript{2}\xspace}%
\newcommand{\ef}{E\textsubscript{F}\xspace}%

\author{Jiaqi~Zhou}
\affiliation{Fert Beijing Institute, BDBC, School of Microelectronics, Beihang University, Beijing 100191, China}
\alsoaffiliation{Centre de Nanosciences et de Nanotechnologies, CNRS, Universit\'{e} Paris-Sud, Universit\'{e} Paris-Saclay, Palaiseau 91120, France}

\author{Junfeng~Qiao}
\affiliation{Fert Beijing Institute, BDBC, School of Microelectronics, Beihang University, Beijing 100191, China}

\author{Chun-Gang~Duan}
\affiliation{Key Laboratory of Polar Materials and Devices, Ministry of Education, East China Normal University, Shanghai 200062, China}

\author{Arnaud~Bournel}
\affiliation{Centre de Nanosciences et de Nanotechnologies, CNRS, Universit\'{e} Paris-Sud, Universit\'{e} Paris-Saclay, Palaiseau 91120, France}

\author{Kang~L.~Wang}
\affiliation{Device Research Laboratory, Department of Electrical Engineering, University of California, Los Angeles 90095, United States}

\author{Weisheng~Zhao}
\email{weisheng.zhao@buaa.edu.cn}
\affiliation{Fert Beijing Institute, BDBC, School of Microelectronics, Beihang University, Beijing 100191, China}

\title[An \textsf{achemso} demo]
  {Large tunneling magnetoresistance in \VSe/\MoS 
  	 magnetic tunnel junction}
\newpage

\begin{document}

\begin{tocentry}

\centering
\includegraphics[height=3.5cm]{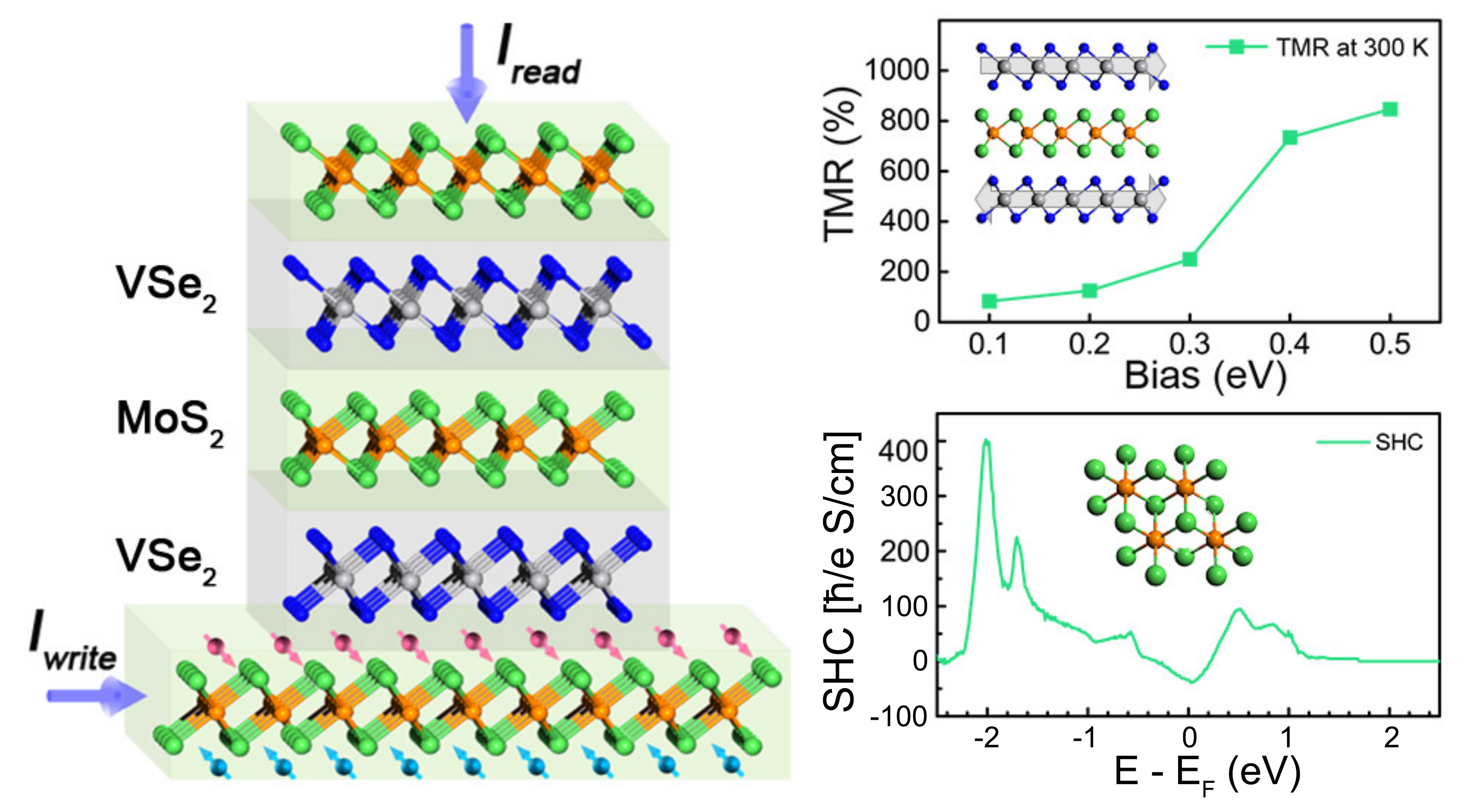}
\label{fig:toc1}

%
%
%

\end{tocentry}

\begin{abstract}
 
Two-dimensional (2D) van der Waals (vdW) materials provide  the possibility of realizing heterostructures with coveted properties. 
Here, we report a theoretical investigation of the  vdW magnetic tunnel junction (MTJ) based on \VSe/\MoS heterojunction, where the \VSe monolayer acts as the  ferromagnet  with the room-temperature ferromagnetism.
We propose the concept of  spin-orbit torque (SOT) vdW MTJ with   reliable reading and efficient  writing   operations.
The non-equilibrium study reveals a large tunneling magnetoresistance (TMR) of 846~\%  at 300 Kelvin, identifying significantly its parallel and anti-parallel states. 
Thanks to the strong spin Hall conductivity of \MoS, SOT is promising for the  magnetization switching of \VSe free layer.
Quantum-well states come into being and resonances appear in MTJ, suggesting that the voltage control can adjust transport properties effectively.
The SOT vdW MTJ based on \VSe/\MoS  provides desirable performance and experimental feasibility, offering new opportunities for 2D spintronics.

\end{abstract}



\newpage

\section{Introduction}

Two-dimensional (2D) van der Waals (vdW) materials, with the strong covalent intralayer bonding and the weak vdW interlayer interaction, exhibit novel properties and various  advantages\cite{Mounet2018Feb,Patel2019,Zhong2017May,Tao2018}. 
Recently, the investigations on ferromagnetic vdW materials make  great progress\cite{Zhao2016,Si2015Aug,Duong2017,Jiang2018} 
and experimental advances have demonstrated that ferromagnetic order can persist in the ultrathin atomic limit\cite{Gong2017Apr,Chen2018}.
Monolayer CrI\textsubscript{3} has been reported to be an Ising ferromagnet with perpendicular magnetic anisotropy (PMA) and 45 Kelvin Curie temperature. \cite{Huang2017Jun} 
After that, Fe\textsubscript{3}GeTe\textsubscript{2}   was demonstrated to possess robust 2D ferromagnetism in monolayer with PMA and 130 Kelvin Curie temperature,\cite{Fei2018Aug} and the ionic gate raises the Curie temperature of Fe\textsubscript{3}GeTe\textsubscript{2} to 300 Kelvin\cite{Deng2018Oct}. 
Moreover, the intrinsic room-temperature ferromagnetism was found in \VSe monolayers\cite{Bonilla2018Feb,Lee2019Jan}. 1T phase VSe\textsubscript{2} monolayers were grown by molecular beam epitaxy (MBE) on \MoS and graphene  substrates, respectively,
and \MoS has been proved to be a good substrate for the uniform growth of ferromagnetic VSe\textsubscript{2} monolayer. 

Ferromagnetic materials are pivotal in the spintronics device, e.g., magnetic tunnel junction (MTJ), which employs the relative magnetization orientation to store binary data.\cite{Wolf2001Nov} MTJ presents high conductance for two ferromagnetic layers in the parallel configuration (PC), while low conductance for two ferromagnetic layers in the anti-parallel configuration (APC), and the difference between the two kinds of conductance is the tunneling magnetoresistance (TMR) effect. MTJs based on vdW materials have been studied a lot.\cite{Cardoso2018Aug,Kim2018Jul,Galbiati2018}. 
Thanks to the perfect spin filtering effect,
high magnetoresistances were predicted in the graphene/Ni (Co) systems through  \textit{ab initio} calculations. \cite{Karpan2007Oct}
The spin valve device using \MoS as the nonmagnetic spacer was prepared, and its magnetoresistance effect was investigated experimentally and theoretically.\cite{Wang2015Aug} 
Magnetic vdW materials were also employed as the ferromagnetic layer in several works. 
        The tunneling spin valve has been  prepared with two exfoliated Fe\textsubscript{3}GeTe\textsubscript{2} crystals and h-BN tunnel layer, and the TMR signal reached up to 160~\% at 4.2 Kelvin.\cite{Wang2018Jun} 
		More intriguingly, the switching of Fe\textsubscript{3}GeTe\textsubscript{2} has been achieved via spin-orbit torque (SOT) induced by spin Hall effect (SHE). \cite{Wang2019Feb,Alghamdi2019Mar}
The vdW heterojunctions consisting of CrI\textsubscript{3} exhibit more superior TMR performances. Layer-dependent magnetic phases have been observed in CrI\textsubscript{3}-based MTJ,\cite{Song2018Jun} and TMR was drastically enhanced with increasing CrI\textsubscript{3} layer thickness, reaching up to 19000~\% at the ultra-low temperature of 2 Kelvin.
Works above reveal the possibility to shrink the MTJ scale to the atomically scale with versatile vdW materials, as well as the TMR enhancement and SOT switching in vdW MTJs.
However, to the best of our knowledge,  the room-temperature TMR remains absent in experimental works of vdW MTJs. Besides, the conventional magnetic field switching hinders the scaling down of vdW MTJ. 

In this work, we design the SOT vdW MTJ based on \VSe/\MoS  heterojunction. By the \textit{ab initio} calculation,  we investigated the intrinsic spin Hall conductivity (SHC) of \MoS, which is the MTJ bottom layer for writing current   injection. The SHC of \MoS is tunable and considerable, 
which suggests that SOT is a promising switching method.
We studied  the spin-resolved transport properties of the SOT vdW MTJ, revealed the quantum-well resonances in vdW layers, and discovered a non-equilibrium TMR up to 846~\% at 300 Kelvin. 
With the merit of experimental feasibility, the SOT vdW MTJ provides efficient writing and reliable reading operations at the room temperature, as well as  practical prospects for spintronics device. 

%

\section{Methods}

The calculations of structure optimization were performed using the density functional theory (DFT) approach implemented in the VASP code,\cite{vasp1}  considering the optB88-vdW function and the Hubbard \textit{U} term. The calculation of SHC was performed by {\sc Quantum ESPRESSO} and WANNIER90  packages.\cite{Giannozzi2009,Marzari2012,Qiao2018Dec} The electrical transport properties   were calculated using OpenMX package through the DFT combined with non-equilibrium Green function (NEGF) formalism considering the Hubbard term.\cite{openmx1}  The generalized gradient approximation (GGA) of Perdew-Burke-Ernzerhof (PBE)\cite{Perdew2008Apr} was used to describe the exchange-correlation functional of the electrons. More details can be found in the Supporting Information. At equilibrium state, the spin-resolved conductance is obtained by the Landauer-B\"{u}ttiker formula
\begin{equation}
G_\sigma=\frac{e^2}{h}\sum_{\boldsymbol{k_{||}}}T_\sigma(\boldsymbol{k_{||}}, E_F),
\end{equation}  
where $T_\sigma(\boldsymbol{k_{||}}, E_F)$ is the transmission coefficient with spin $\sigma$ at the transverse Bloch wave vector $\boldsymbol{k_{||}}$ and  Fermi energy $E_F$, $e$ is the electron charge and $h$ is the Planck constant. At non-equilibrium state, the current is calculated as the integral
\begin{equation}
\label{current}
I_\sigma=\frac{e}{h}\int T_\sigma(E)[f(E-\mu_1)-f(E-\mu_2)]dE,
\end{equation}  
where $f$ is the Fermi distribution function, $\mu_1$ and  $\mu_2$ are the chemical potentials of the left and right leads, respectively. 

\section{Results and discussions}

\begin{figure}[!htb]
	\centering
	\includegraphics{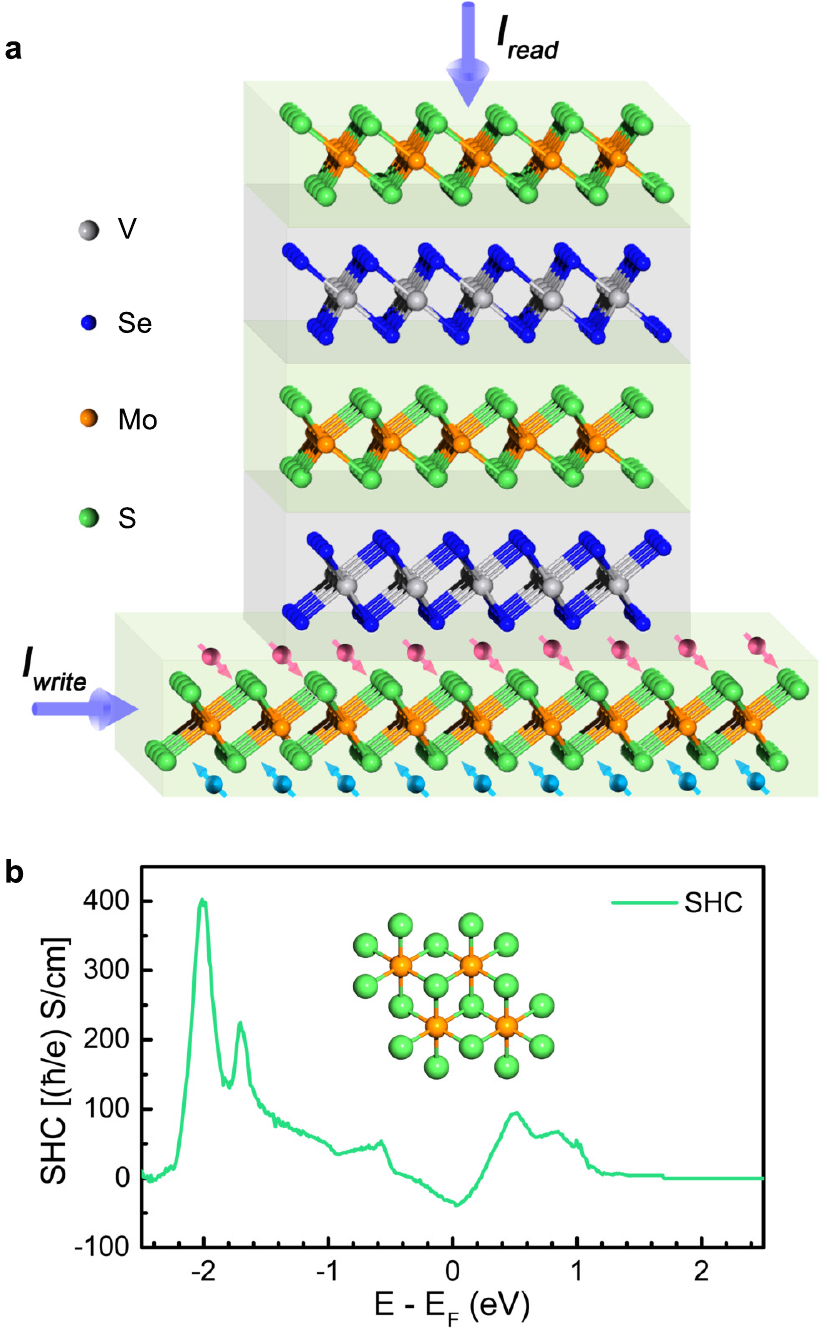}
	\caption{The concept of SOT vdW MTJ. \textbf{a} The atomic schematic diagram of the SOT vdW MTJ. The stack from  bottom to top is 1T \MoS/1T \VSe/1H \MoS/1T \VSe/1T \MoS.		
		The writing operation uses the in-plane current in the bottom \MoS layer, and the reading operation employs the TMR effect in the vertical junction. Arrows in purple indicate the writing and reading currents, and the small arrows in pink and cyan represent the spins of electrons. \textbf{b} SHC of the bulk 1T \MoS material. The directions of charge current, spin current and spin is perpendicular to each other, and the spin current is in the vertical direction.  }
	\label{fig:sot}
\end{figure}

The atomic schematic diagram of the SOT vdW MTJ is shown in Figure \ref{fig:sot}\textbf{a}. 
The \VSe  in 1T phase is employed as the ferromagnetic layer. As the magnetization of \VSe decays with the  increasing layer number,\cite{Shabbir2018Dec} we only studied the monolayer \VSe in this work. 
1H phase \MoS is used as the tunnel barrier layer with the band gap of 1.9 eV.\cite{Zeng2012Jun} 
\MoS is a good substrate for uniform growth of \VSe,\cite{Bonilla2018Feb} thus the \VSe/\MoS heterojunction has high feasibility in experiments.
	Apart from the basic MTJ structure, the metallic vdW material,  1T phase \MoS  is adopted as the electrode in the top and bottom of MTJ. 
		Thanks to the advances of synthesis technology,\cite{Liu2015Nov,Sharma2018Aug} 		metallic 1T phase MoS\textsubscript{2} nanosheets can be stabilized and  employed as the electrode. \cite{Acerce2015Mar}
The SOT vdW MTJ is symmetrical to keep the most stable interfacial configurations, which have been confirmed by VASP calculations. Details about the optimized structure can be found in Figure S1 in Supporting Information.

The basic manipulations of MTJ include reading and writing operations. Several mechanisms can be employed in the writing process, such as  the external magnetic field,\cite{Wang2018Jun}  the spin transfer torque (STT),\cite{Wang2018Feb} as well as  SOT. \cite{Wang2019Feb}
The  magnetic field switching impedes the device miniaturization,  while the STT is not feasible due to the low spin polarization of 1T \VSe. \cite{Ma2012Jan}
		With the merits of simple device structure, high reliability, and remarkable efficiency, we choose SOT switching in the vdW MTJ. 
		One of  SOT mechanisms is  spin Hall effect (SHE), which is a phenomenon that spin-orbit coupling effect generates an asymmetric deviation of the charge carriers due to the different spin direction\cite{RevModPhys.87.1213}. SHE realizes the  conversion from charge current to spin current. 
		In the bilayer system, the ferromagnetic layer would absorb the  angular momentum  of the spin current in nonmagnetic layer, and achieve SOT   switching of magnetization in ferromagnetic layer. \cite{Manchon2018Jan,Brink2016Mar,Khang2018Jul}
		SOT switching of  Fe\textsubscript{3}GeTe\textsubscript{2}  magnetization has been observed in the  Fe\textsubscript{3}GeTe\textsubscript{2}/Pt bilayer\cite{Wang2019Feb,Alghamdi2019Mar}. 
		SHE in Pt produces a pure spin current, which enters the Fe\textsubscript{3}GeTe\textsubscript{2} layer and induces both field-like and damping-like torques.\cite{Alghamdi2019Mar}  Compared to the conventional SOT devices, high SOT efficiency has been found in the Fe\textsubscript{3}GeTe\textsubscript{2}/Pt bilayer.
		The efficiency enhancement could be  attributed to the atomically flat surface, which may conduce to the proximity effect in vdW heterostructures.\cite{Safeer2019Jan} Discussions above illustrate that vdW MTJs 
		have the potential to achieve SOT switching via SHE. 
As Figure \ref{fig:sot}\textbf{a} shows, when the writing current is injected into the bottom \MoS electrode, the spin current arises in the vertical direction with the in-plane spin torque, which can be used to switch the magnetization of the \VSe layer adjoining to the bottom \MoS layer. 
To explore the strength of SHE,  
 we studied the intrinsic spin Hall conductivity (SHC) of 1T \MoS.
Results are shown in Figure \ref{fig:sot}\textbf{b}. Although the SHC at \ef is limited to  $-$34 {($\hbar$/e)S/cm}, SHC at E = $-$2.0 eV reaches up to 400 {($\hbar$/e)S/cm}. It indicates that gate voltage control is an efficient manipulation  to improve SHC. As the SHC in 1T \MoS is comparable to that in Weyl semimetals with strong SOC\cite{Zhou2019Feb}, the MTJ based on \VSe/\MoS heterojunction is promising to be switched by the SOT effect. 



Apart from the writing operation, another critical manipulation in MTJ is the reading process, which relies on the TMR effect. High TMR is a desirable performance for the reliability of MTJ device, and TMR at room temperature is essential to the practical prospect. We studied the TMR in the above-mentioned MTJ, abbreviated as MT MTJ due to the 1H \MoS tunnel barrier layer. 
For MT MTJ, we define the PC (APC) as the bottom \VSe monolayer having the parallel (anti-parallel) magnetic orientation relative to the top \VSe monolayer, as the bottom \VSe monolayer is the free layer. 
Besides, we design a new vdW MTJs with 1H \VSe as the tunnel barrier layer, abbreviated as VT MTJ.   
For VT MTJ, we define the PC (APC) as the 1H \VSe tunnel layer holding the parallel (anti-parallel) magnetic orientation to two 1T \VSe monolayers, namely, 1H \VSe is the free layer. 
1H  \VSe is the magnetic semiconductor,\cite{Tong2016Dec} so it is predictable that both tunnel transport and spin filtering would happen in the VT MTJ, as well as high TMR. 
	The interlayer exchange coupling (IEC) in VT MTJ was investigated. It turns out that IEC in VT MTJ is comparable with Fe/MgO/Fe system, indicating the data storage stabilization of  VT MTJ. 
	The possibility of VT MTJ switching by anomalous Hall effect was also considered.  
Details about  IEC analysis and VT MTJ switching can be found in Supporting Information.

\begin{table} [b]
	\centering 
	\caption{ Spin-resolved conductance and TMR at the equilibrium state in MT MTJ and VT MTJ. The conductance is at 300 Kelvin  in the unit of  $10^{-5}$ $e^2/h$.}
	\label{tab:TMR}
	\begin{tabular}{cccccc} 
		\hline
		MTJs &$G^{\uparrow}_{PC}$ &$G^{\downarrow}_{PC}$ &$G^{\uparrow}_{APC}$ &$G^{\downarrow}_{APC}$ &TMR(\%)\\ 
		\hline 
		MT MTJ   &225  &45  &111  &111  & 22 \\
		VT MTJ   &69   &182 &34   &9    &484 \\
		\hline
	\end{tabular}
\end{table}

We calculated the spin-resolved conductance and TMR of both MT and VT MTJs at the equilibrium state at 300 Kelvin, and present the results in Table \ref{tab:TMR}. 
The TMR is defined as $TMR = \frac{G_{PC}-G_{APC}}{G_{APC}}\times100~\%$ at equilibrium state, where $G_{PC}$ and $G_{APC}$ is the total conductance for the PC and APC of MTJ, respectively.  $G_{PC} = G^{\uparrow}_{PC}+G^{\downarrow}_{PC}$, where $G^{\uparrow}_{PC}$ and $G^{\downarrow}_{PC}$ is the majority-spin  and minority-spin conductance in PC, respectively.
$G_{APC} = G^{\uparrow}_{APC}+G^{\downarrow}_{APC}$, where $G^{\uparrow}_{APC}$ and $G^{\downarrow}_{APC}$ is the majority-spin and minority-spin conductance in APC, respectively. 
The TMR in MT MTJ is only 22~\%, however, the VT MTJ presents a much remarkable TMR up to 484~\%.
 In both MTJs, the conductances have the comparable scale in PC while those in APC vary a lot. 
 The $G^{\uparrow}_{APC}$ in MT MTJ is three times as large as that in VT MTJ, and more distinctly, the $G^{\downarrow}_{APC}$ in MT MTJ is over one magnitude larger than that in VT MTJ. 
 As the TMR is inversely proportional to $G_{APC}$, the low $G_{APC}$ results in high TMR in VT MTJ. We attribute the high TMR  to the efficient spin filtering effect in APC of   VT MTJ, where \VSe trilayer is stacked in anti-ferromagnetic ordering, blocking spin transport  especially for the minority spin in APC. In contrast, the high conductance in APC of MT MTJ damages TMR. 
%

\begin{figure} [!htb]
	\centering
	\includegraphics{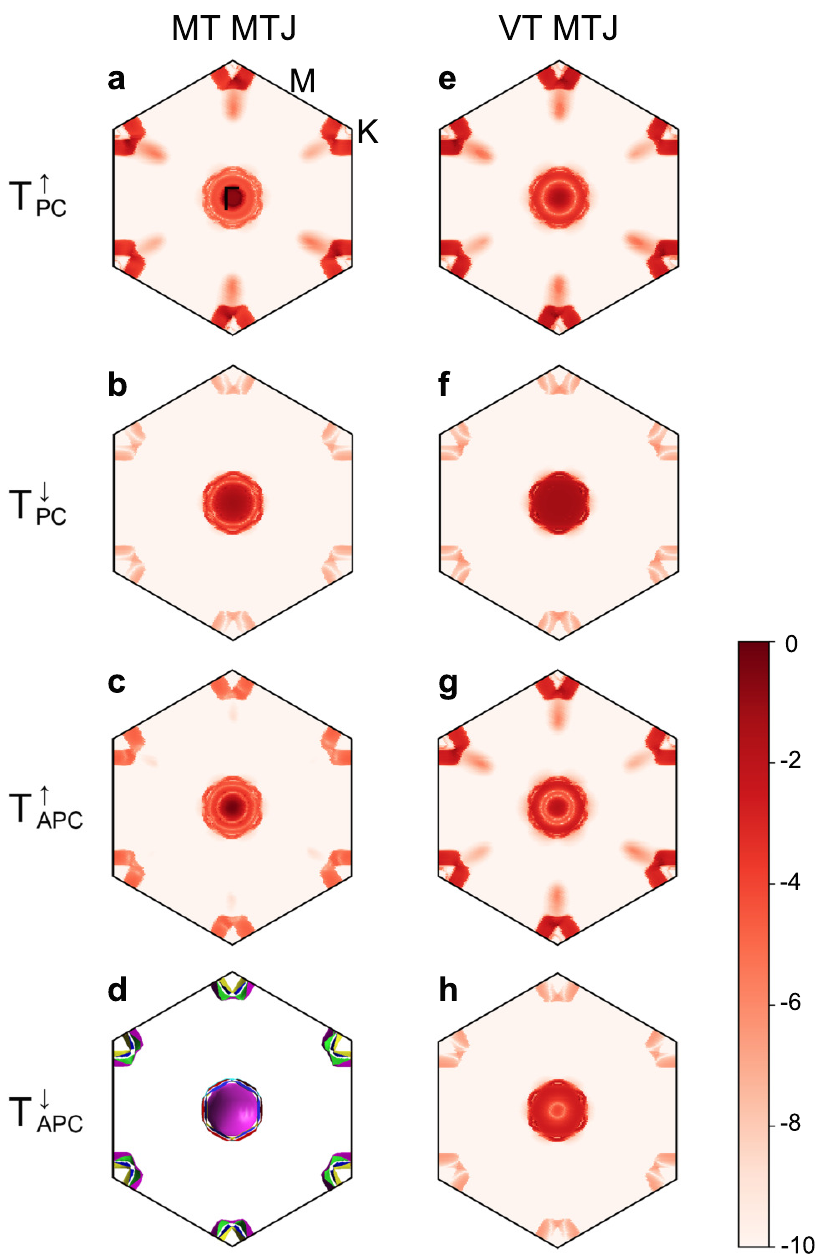}
	\caption{Spin- and $\boldsymbol{k_{||}}$-resolved transmission spectra at equilibrium states  in log scale. 
 \textbf{a}-\textbf{c} are for MT MTJs and \textbf{e}-\textbf{h} are for VT MTJ. 
	$T^{\uparrow}_{PC}$ is for majority-spin channel and $T^{\downarrow}_{PC}$ is for minority-spin channel in the parallel configuration. 
	$T^{\uparrow}_{APC}$ is for majority-spin channel and $T^{\downarrow}_{APC}$ is for   minority-spin channel in the anti-parallel configuration. 
	Note that in MT MTJ, $T^{\uparrow}_{APC}$ is the same as $T^{\downarrow}_{APC}$, and we only show the  $T^{\uparrow}_{APC}$ in \textbf{c}. 
	\textbf{d} The Fermi surface of 1T \MoS electrode.  	
	The color bar exhibits the transmission scale and deep red color represents high transmission. }
	\label{fig:trans}
\end{figure}

Although the VT MTJ can realize high TMR at equilibrium state, it is not feasible for the SOT vdW MTJ, as the free layer of VT MTJ is in the middle of the device, not adjoin the bottom layer. We may make the bottom 1T \VSe ferromagnetic layer as the free layer, but the double spin filtering effect would vanish, further impair TMR. Consequently, it is imperative to analyze the factor leading to low TMR and improve it in MT MTJ. 
According to the  Landauer-B\"{u}ttiker formula, the conductance is the integration of transmission over all $\boldsymbol{k_{||}}$ points in the 2D  Brillouin zone (BZ). We plot the $\boldsymbol{k_{||}}$-resolved transmission spectra to present more details in the BZ. 
Figure \ref{fig:trans}\textbf{a}-\textbf{h} show the transmission spectra in different spin channels in both MT and VT MTJs except for Figure \ref{fig:trans}\textbf{d}, which is the Fermi surface of 1T \MoS electrode, determining the outlines of the transmission spectra. 
In $T^{\uparrow}_{PC}$ of MT MTJ, a sharp peak  with the $T^{\uparrow}_{PC} = 0.16$ at $\Gamma$ point results in the high conductance in $G^{\uparrow}_{PC}$. Due to the symmetrical structure of MT MTJ, the transmission spectra of $T^{\uparrow}_{APC}$ and $T^{\downarrow}_{APC}$ is the same at equilibrium state, so we only show the $T^{\uparrow}_{APC}$ in Figure  \ref{fig:trans}\textbf{c}. A sharp peak arises around $\Gamma$ point as shown by the deep red color in Figure  \ref{fig:trans}\textbf{c}, inducing in the high conductance  $G^{\uparrow}_{APC} = G^{\downarrow}_{APC} = 1.11\times10^{-3}$ $e^2/h$ in MT MTJ. We find that $T^{\uparrow}_{APC}= T^{\downarrow}_{APC} = 0.57 $ at $\Gamma$ point, an extraordinary transmission coefficient for the tunneling transport. 
On the other hand, for VT MTJ, a remarkable  broad peak around $\Gamma$ point is observed in $T^{\downarrow}_{PC}$, and the transmission coefficient $T^{\downarrow}_{PC} = 0.06$ is considerable at $\Gamma$ point, resulting in the high conductance $G^{\downarrow}_{PC}= 1.82\times10^{-3}$ $e^2/h$  in VT MTJ. 
Due to the double spin filtering effect in VT MTJ, the transmission in minority-spin channel $T^{\downarrow}_{APC}$ in Figure  \ref{fig:trans}\textbf{h} is much weaker, as low as  $T^{\downarrow}_{APC} = 1.19\times10^{-4}$ at $\Gamma$ point. 

\begin{figure} [!htb]
	\includegraphics[width=17cm]{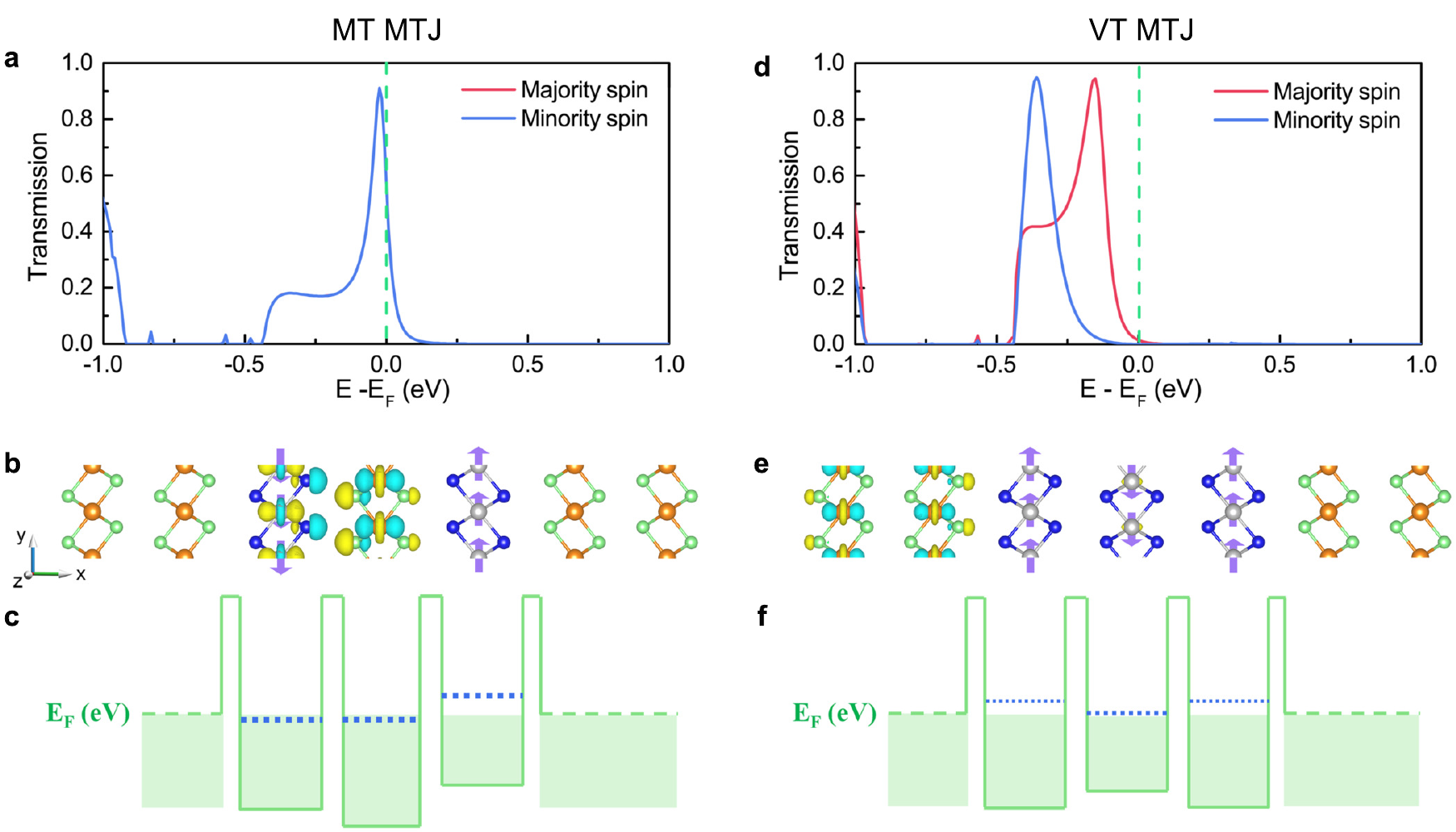}
	\caption{Transmission analysis of $T^{\downarrow}_{APC}$ at $\Gamma$ point for MT and VT MTJs, and the left column is for MT MTJ while the right column is for VT MTJ. 
	\textbf{a} and \textbf{d} are transmission coefficients versus energy. The red color indicates the majority-spin transmission while the blue line indicates the minority-spin transmission. In MT MTJ, the red line is overlapped by the blue line and invisible.
	\textbf{b} and \textbf{e} are transmission eigenchannel wavefunctions. 	
	Small arrows show the magnetizations of \VSe. \textbf{c} and \textbf{f} are schematic diagrams of the potential profiles. The blue dotted line indicates the minority-spin quantum-well states.  The green dashed lines present the location of \ef, set as zero.	  	 }
	\label{fig:diagram}
\end{figure}

To understand the unusual transmission coefficient and clarify the APC transmission, we present the minority-spin transmission analysis at  $\Gamma$ point for APC of both MT and VT MTJs, where $T^{\downarrow}_{APC} = 0.57$ and $1.19\times10^{-4}$, respectively. The analysis of components of wavefunctions indicates that the \textit{s} and \textit{p\textsubscript{x}} orbitals in S and Se atoms, as well as the  \textit{s}, \textit{p\textsubscript{x}}, \textit{d\textsubscript{x\textsuperscript{2}-y\textsuperscript{2}}} and \textit{d\textsubscript{z\textsuperscript{2}}} orbitals in Mo and V atoms, make the dominant contribution to the transmission, especially \textit{d} orbitals. The shapes of \textit{p\textsubscript{x}} orbitals in S and Se atoms can be observed by the wavefunctions in Figure \ref{fig:diagram}\textbf{b} and \textbf{d}, as well as the \textit{d\textsubscript{z\textsuperscript{2}}} orbitals in Mo and V atoms.
Figure \ref{fig:diagram}\textbf{a} shows  the transmission versus energy curve in MT MTJ. The transmission peaks of both spins locate approximately at E\textsubscript{F}, at E =  $-$0.02 eV to be exact. The transmission coefficient at \ef is $T^{\downarrow}_{APC} = 0.57$, which is extraordinarily high for tunneling transport.  In Figure \ref{fig:diagram}\textbf{b}, it is distinctive that eigenchannel wavefunctions localize at the left \VSe and the middle \MoS tunnel layers. Because of the weak interaction among vdW layers and the strong intralayer covalent bond, quantum-well resonances are very likely to arise in vdW heterojunction\cite{Koleini2007May,Song2018Jun}. 
 Considering the high transmission and wavefunction localization, we can infer that the quantum-well resonances arise in MT MTJ.  
 		Quantum-well states exhibit the localization of the transmission eigenchannel wavefunctions. The electrons are reflected back and forth in the barrier region, forming a standing wave pattern, and results in the almost perfect $\Gamma$ transmission.
Figure \ref{fig:diagram}\textbf{c} gives a more intelligible picture on resonances by the schematic plot of the potential profile. We use blue dot lines in vdW layers to indicate the quantum-well states of the minority spin. The quantum-well states in  the left \VSe and the middle \MoS tunnel layers are very close to the \ef. The match of two  quantum-well states results in the resonances of wavefunctions in this area, 
as well as the intense localization and  remarkable transmission. 
As the quantum well states are sensitive to the location of \ef, we studied the transmission with \ef shift and present the results in Figure S2. 
Stronger resonances can be observed at the energy point E =  $-$0.02 eV where transmission  $T^{\uparrow}_{APC} = T^{\downarrow}_{APC} = 0.91$, an extremely high coefficient close to 1. 

Figure \ref{fig:diagram}\textbf{d} shows the transmission  for VT MTJ. Two peaks deviate from the \ef.
$T^{\uparrow}_{APC} = 0.95$  at E =  $-$0.15 eV, and $T^{\downarrow}_{APC} = 0.94$  at E =  $-$0.36 eV. 
 Due to the sharp  attenuation of transmission, the transmissions at E\textsubscript{F} for both spins are inconspicuous, resulting in  $T^{\downarrow}_{APC} = 1.19\times10^{-4}$ at \ef. 
Figure \ref{fig:diagram}\textbf{e} shows the minority-spin eigenchannel wavefunctions for VT MTJ, the wavefunctions decay along the transport x direction from the incoming \MoS electrode. Note that for minority spin, both the left 1T \VSe and the right 1T \VSe   has the prohibitive magnetization, this spin filtering effect results in the fast decay in VT MTJ.  
We plot the schematic of the potential profile for the model of magnetic quantum wells in VT MTJ, as shown in Figure \ref{fig:diagram}\textbf{f}. Due to the energy mismatch among quantum-well  states, no resonance happens and the wavefunctions attenuate very fast along the transport direction. 
However, the transmission peak at  E~=~$-$0.15 eV for majority spin and the peak at  E~=~$-$0.36 eV for minority spin can be explained by the resonance transmission, as shown in Figure S3. 
Investigations above demonstrate that the vdW heterojunction transport properties are sensitive to the quantum-well resonances, and the Fermi energy shifting can be a practical approach to modulate the transmission of vdW heterojunctions. 
		Note the quantum-well states play a critical role in the high TMR, it is experimentally suggested to keep clean interfaces in the vdW heterojunctions, in order to maintain high TMR and avoid the short-circuiting current.
		In addition, we present the analysis of transmission at $\boldsymbol{k_{||}} = (0.283, 0.283)$, where high transmission coefficients are observed. The spin-resolved wavefunctions are shown in Fig. S4. Different from the quantum-well states at  $\Gamma$ point,  wavefunctions at $\boldsymbol{k_{||}} = (0.283, 0.283)$ decay along the transport direction.

\begin{figure} [!htb]
	\centering
	\includegraphics{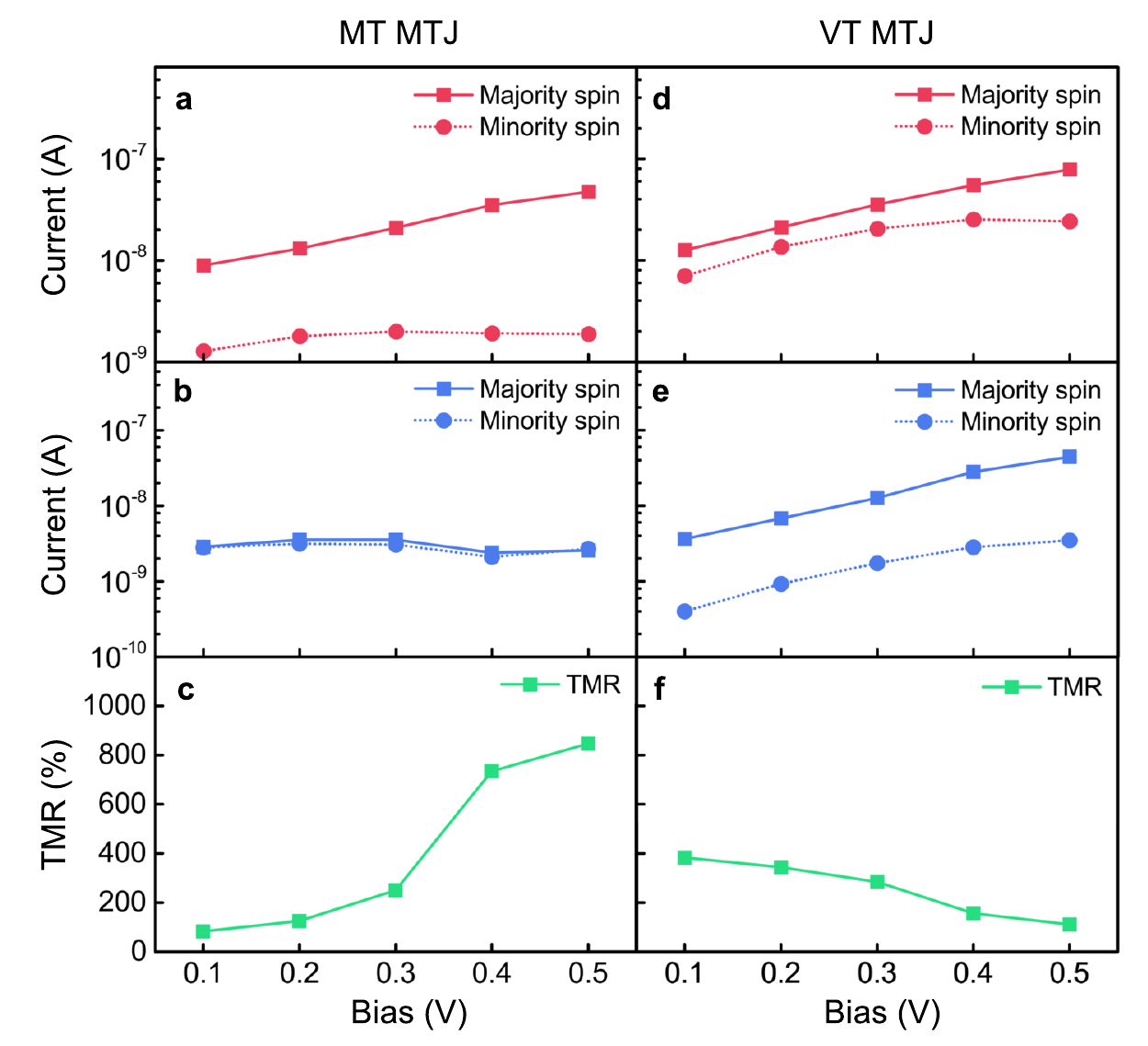}
	\caption{Non-equilibrium transport properties at 300 Kelvin. The left column is for MT MTJ while the right column is for VT MTJ.  \textbf{a} and \textbf{d} are the currents in PC. \textbf{b} and \textbf{e} are the currents in APC. \textbf{c} and \textbf{f} are the TMRs. }
	\label{fig:current}
\end{figure}

Discussions above illuminate that the quantum-well resonances play an important role in the vdW MTJ at equilibrium state. However, non-equilibrium state tells another story. According to Equation \ref{current}, at non-equilibrium state, the current is the integral of transmission with respect to energies in the bias window, rather than the direct result of transmission at \ef. In this condition, the quantum-well resonances have limited influence on the non-equilibrium transmission. By observing the APC transmission spectrum of MT MTJ shown in Figure \ref{fig:diagram}\textbf{a}, we found that the states at energies above \ef contribute little to the transmission. To reduce the current in APC and get a high TMR, we applied positive bias to the MT MTJ and calculated the non-equilibrium transport properties. As the reference, non-equilibrium transport in VT MTJ was also studied. 
Figure \ref{fig:current} shows the spin-resolved current and TMR under varying bias. Here the TMR is defined as $TMR = \frac{I_{PC}-I_{APC}}{I_{APC}} \times 100~\%$. We observe that in both MTJs PC currents rise with increasing bias, and majority-spin current is larger than minority-spin current, as shown in Figure \ref{fig:current}\textbf{a} and \textbf{d}. More details on current in BZ can be found in Figure S5 and S6.
However, APC currents behave distinctly  in MT and VT MTJs. 
Figure \ref{fig:current}\textbf{b} shows that in MT MTJ, both majority-spin and minority-spin currents in APC remain almost unchanged with increasing bias, while APC currents in VT MTJ  rise with increasing bias, as shown in Figure \ref{fig:current}\textbf{e}. 
Figure S5 and S6 present more details on this difference from the aspect of BZ. 
In MT MTJ, 
even if the transmission broadens in the center of BZ, the transmission peak at $\Gamma$ point weakens a lot with increasing bias. As a result of trade-off, the current in APC almost remains unchanged with bias in APC. 
In VT MTJ, the current of majority spin around K point strengthens with increasing bias,  and the current of minority spin broadens around $\Gamma$ point, resulting in the APC current enhancement. 
TMR is inversely proportional to the APC current.
 Consequently, in MT MTJ, TMR  augments with the increasing bias because of the changeless APC current, while   TMR in VT MTJ declines due to the enhancement of APC current with increasing bias.  TMR of 846~\% can be observed at 0.5 V bias in MT MTJ at 300 Kelvin, a considerable result for the room-temperature application. 
Apart from the calculation at 300 Kelvin, we present the calculation at 100 Kelvin in Fig. S7. It worth stressing that a TMR up to 1277~\% is observed in MT MTJ, which is attributed to the decline of APC current at 100 Kelvin. This result provides the potential for TMR improvement by the reduction of temperature.

\begin{figure} [!htb]
	\centering
	\includegraphics{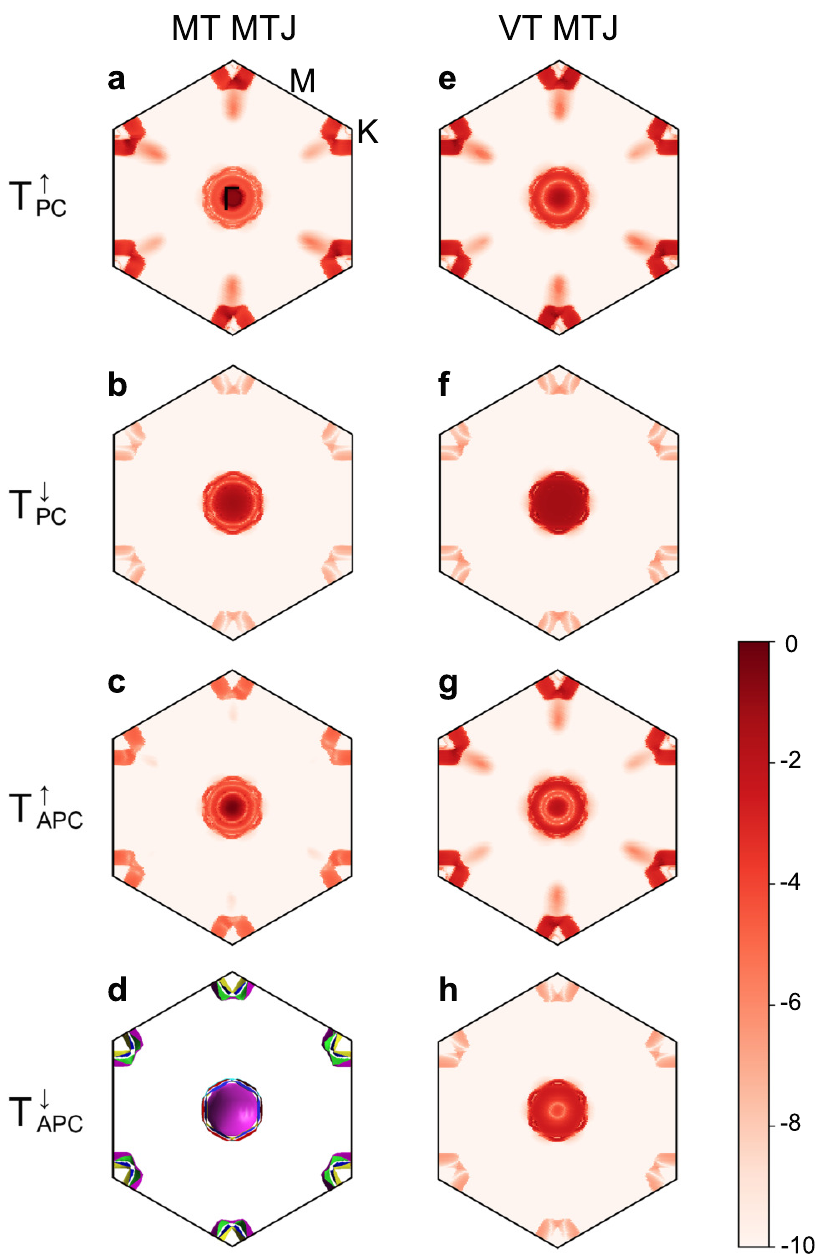}
	\caption{Transmission coefficients in APC at 300 Kelvin 
		integrating over all $\boldsymbol{k_{||}}$ points.  The left column is for MT MTJ while the right column is for VT MTJ.   \textbf{a} and \textbf{d} for 0.1 V bias. \textbf{b} and \textbf{e} for 0.3 V bias. \textbf{c} and \textbf{f} for 0.5 V bias. Bias windows are shown by green dashed lines. }
	\label{fig:transE}
\end{figure}

As the current is obtained by integrating transmission with respect to energies in the bias window, we now turn to the analysis of the transmission  versus energy at  finite bias.  
The above analysis  illustrates that  the APC transmission leads to the difference of TMR in MT and VT MTJs, so we present the result of transmission versus energy for APC, and the transmission  is the integration over all $\boldsymbol{k_{||}}$ points in the BZ, as shown in Figure \ref{fig:transE}. 
It can be observed that  the transmission spectra for MT MTJ are almost changeless in the bias window for any bias, and the transmission coefficients are approximately zero. 
The $\Gamma$  point resonances of quantum-well states at E =  $-$0.02 eV  make the electron intensely localized. Under the influence of bias, the resonances shift to the negative energy,  at E =  $-$0.08 eV for the 0.5 V bias, out of the bias windows in the non-equilibrium transmission, and result in a suppressed current in APC. 
On the other hand, in VT MTJ transmission, majority spin contributes more than minority one. 
With the  rising bias, more majority-spin transmission peaks enter into the bias window and contribute to the current.
Due to the double spin filtering layer for minority spin, the transmission is suppressed close to zero over all the bias windows. 
Combining the majority-spin and minority-spin current, the total current in APC increases with bias in VT MTJ. 
Figure S8 presents  the transmission  versus energy at 100 Kelvin.    It   has a similar outline with the transmission spectra at 300 Kelvin. Note that the decreasing temperature declines the conductance in APC of MT MTJ, and results in the TMR  enhancement.
The analysis above explains the differences between APC currents in MT and VT MTJs, and clarifies the reason for distinct TMR behaviors.

\section{Conclusion}
In conclusion, we propose an MTJ based on vdW heterojunction  consisting of 2D ferromagnet \VSe and \MoS monolayer, and validate its performances by \textit{ab initio} calculation.
A large TMR at room temperature over 800~\%  can be  obtained  through the voltage control, enhancing the sensitivity of MTJ. 
Thanks to the considerable and tunable  SHC of \MoS, the MTJ is promising to be switched efficiently by SOT.
 It is important to mention that
we fully consider the further practical implementation, as the \VSe monolayer with room-temperature ferromagnetism has been demonstrated experimentally on the \MoS substrate.
This SOT vdW MTJ  offers new prospects for low-dimensional spintronics applications.

\begin{acknowledgement}

The authors thank the National Natural Science 
Foundation of China (Grant No. 61627813, 61571023), the International 
Collaboration Project B16001, and the National Key Technology Program 
of China 2017ZX01032101 for their financial support of this work.
 This work is supported by the Academic Excellence Foundation of BUAA for PhD Students. The calculations were performed on TianHe-1A supercomputer at National Supercomputer Center in Tianjin. 

\end{acknowledgement}

\begin{suppinfo}

%
\begin{itemize}
  \item Supp\_info.pdf: Statement on \textit{ab initio} calculation methods,  VT MTJ properties, optimized structures, eigenchannel wavefunctions,  wavefunctions at K' point, current in BZ, and NEGF transport properties at 100 Kelvin. 
\end{itemize}

\end{suppinfo}


\providecommand{\latin}[1]{#1}
\makeatletter
\providecommand{\doi}
{\begingroup\let\do\@makeother\dospecials
	\catcode`\{=1 \catcode`\}=2 \doi@aux}
\providecommand{\doi@aux}[1]{\endgroup\texttt{#1}}
\makeatother
\providecommand*\mcitethebibliography{\thebibliography}
\csname @ifundefined\endcsname{endmcitethebibliography}
{\let\endmcitethebibliography\endthebibliography}{}

\end{document}